\documentclass[a4paper,11pt]{article}
\pdfoutput=1
\usepackage{amsfonts,latexsym,graphicx,amssymb, amsmath}

\pdfoutput=1

\newcommand{\beq}{\begin{equation}}
\newcommand{\eeq}{\end{equation}}
\newcommand{\beqa}{\begin{eqnarray}}
\newcommand{\eeqa}{\end{eqnarray}}
\newcommand{\beql}{\begin{align}}
\newcommand{\eeql}{\end{align}}

\newcommand{\pd}{\partial}
\newcommand{\nn}{\nonumber}

\newcommand{\R}{\mathbb{R}}

\newcommand{\lp}{\left(}
\newcommand{\rp}{\right)}

\DeclareMathOperator\arctanh{arctanh}

\usepackage{jheppub}
\usepackage{graphicx}
\usepackage{subcaption}
\usepackage{dsfont}
\usepackage[dvipsnames]{xcolor}
\usepackage{etoolbox}
\usepackage[T1]{fontenc}
\usepackage{amsmath}
\usepackage{mathtools}
\usepackage{amsthm}
\usepackage{physics}
\makeatletter
\patchcmd{\maketitle}{\@fpheader}{\\}{}{}
\usepackage{stackengine}
\stackMath

\makeatother

\title{Holography of time machines}
\author[a,b]{Roberto Emparan}
\author[b]{and Marija Toma\v{s}evi\'c}

\affiliation[a]{Instituci\'o Catalana de Recerca i Estudis Avan\c cats (ICREA),
Passeig Llu\'{\i}s Companys 23, E-08010 Barcelona, Spain}
\affiliation[b]{Departament de F{\'\i}sica Qu\`antica i Astrof\'{\i}sica, Institut de Ci\`encies del Cosmos,
Universitat de Barcelona, Mart\'{\i} i Franqu\`es 1, E-08028 Barcelona, Spain}

\emailAdd{emparan@ub.edu}
\emailAdd{mtomasevic@icc.ub.edu}

\abstract{We use holography to examine the response of interacting quantum fields to the appearance of closed timelike curves in a dynamically evolving background that initially does not contain them. For this purpose, we study a family of two-dimensional spacetimes that model very broad classes of wormhole time machines. The behavior of strongly coupled conformal theories in these spacetimes is then holographically described by three-dimensional AdS bulk geometries that we explicitly construct. The dual bulk spacetime is free from any divergences, but splits into two disconnected components, without and with CTCs, which are joined only through the boundary; then, passages across the chronology horizon are impossible for any field excitations. In dual terms, the strong self-interaction of the CFT decouples the pathological part from the rest of the spacetime. We also find that entangling the CFTs in two separate time machines connects them through a traversable bulk wormhole. Nevertheless, any entanglement-assisted chronology violations will be prevented by quantum bulk corrections, i.e., subleading $1/N$ effects, again without needing any gravitational backreaction of the CFT. We are led to speculate that chronology may be protected without involving Planck scale physics.
}

\begin{document}

\maketitle

\section{Introduction}
\label{intro}

Time machines, or spacetimes with closed timelike curves (CTCs), are infamous for their often unpleasant and paradoxical consequences. Yet, it has never been fully clarified whether they necessarily contradict the known laws of physics. Various no-go theorems exist which, under physically plausible assumptions, deem the operation of a time machine impossible. For instance, the achronal average null energy condition excludes the possibility of time machines simply by incompatibility \cite{Graham_2007}. Nevertheless, studying spacetimes that develop CTCs can uncover certain aspects of the laws of physics that are not manifest otherwise.

In this paper, we will employ the AdS/CFT correspondence to examine the response of quantum fields to the creation of a time machine.
To be clear, our goal is not to show whether holographic quantum gravity forbids CTCs to arise in a spacetime. Instead, we will apply AdS/CFT to investigate how self-interaction affects the propagation of quantum fields in simple models of such spacetimes. Our work then continues along the early studies \cite{Hiscock:1982vq,PhysRevD.46.603, Morris:1988tu, Klinkhammer:1992tb, Boulware:1992pm, Visser:1995cc, PhysRevD.43.3878} that showed that the stress tensor of a free quantum field theory develops a divergence in the approach to the chronology (Cauchy) horizon.\footnote{The chronology horizon exists in spacetimes in which we have localized creation of CTCs. Spacetimes that have CTCs at every point, and thus no chronology horizon, were studied in a holographic setup in \cite{Arefeva_2016}. } This restriction to free fields is of course technically convenient, but self-interactions could be important even before the divergence develops. Nevertheless, little is known about these effects.\footnote{Ref.~\cite{Hiscock:2000jq} treats interactions only perturbatively, which, perhaps unsurprisingly, almost does not modify the picture.}

The AdS/CFT correspondence not only provides the means to solve the strongly coupled theory and verify that essentially the same divergence of the stress tensor is still present; it also maps the question of what happens when a quantum excitation attempts to cross the chronology horizon, into an analysis of the geometric properties of the bulk. This perspective is, we believe, useful. Note that, in a fixed time-machine geometry, nothing prevents a free quantum field excitation from crossing to the side with CTCs: even though the field may be in a state that is singular on the horizon, we can still create localized wavepackets, on top of that state, which travel along geodesics across the horizon. Instead, a self-interacting field excitation is sensitive to the background field state. For a holographic CFT, this is described by a particle following a geodesic in the bulk. However, a divergence in the holographic stress tensor only implies a divergence in the extrinsic curvature of certain hypersurfaces in the bulk, and this need not be a bulk curvature singularity; indeed it cannot be, if the bulk is (as in our constructions) locally a constant curvature AdS geometry. So, if the bulk geometry is not singular, what can possibly prevent a quantum excitation of the field from crossing the chronology horizon? 

Eventually, a large gravitational backreaction from the diverging stress tensor is expected. But even if the gravitational backreaction creates a singularity, it is not obvious that it will prevent the passage of all excitations to the CTC side. We may envisage, for instance, a holographic setup where the boundary geometry is dynamical (e.g., as in braneworld holography), and backreaction creates a bulk curvature singularity, but one that does not extend throughout all of the bulk spacetime but might be bypassed by diving deep enough in it (in dual terms, this would be a very long wavelength quantum that would not sense the singularity). To see whether this happens or not, a proper calculation of the backreaction would be needed. 
Still, we find that, even with gravitational dynamics turned off, excitations of a self-interacting field can never cross the chronology horizon. This suggests that protection of the normal time order may be a very general consequence of quantum effects. 

This study of quantum field propagation in a potentially pathological fixed curved background can also be put in a broader context. The stress-energy tensor of quantum fields is expected to be divergent on other Cauchy horizons, sometimes not associated to CTCs, as in some black hole interiors. Are there universal patterns to these divergences, which may be captured by universal local models of bulk geometries dual to these divergent stress tensors? While we do not have full answers yet, our study provides a step in this direction: the mechanism that cuts off the region beyond the chronological horizon is remarkably simple and may well be present in other Cauchy horizons.

To investigate these questions, we will consider a broad class of generic two-dimensional geometries with CTCs (Misner spacetimes), and construct the bulk that is dual to the conformal fields in such spacetimes. The boundary spacetime is chosen to exhibit the known features of time machines, and we will reproduce the aforementioned stress tensor divergence. The novelty lies in the peculiar way that the bulk implements the impossibility to cross the chronology horizon. We will show that the bulk fragments into two disconnected components, which are joined only through the boundary. The location of the chronology horizon will then sit on the boundary of the spacetime, making any bulk passages impossible. In addition, we will show that the CTC-region in the bulk becomes geodesically incomplete for all states of the boundary time machine. That is, while in the non-CTC region the CFT is well defined up until the moment when the chronology horizon is reached, in the region with CTCs the CFT fails to be regular in a global way, not restricted to the localized divergences at the chronology horizons.

Our study also reveals an intriguing possibility for chronology violation which, again, ends up failing owing to self-interaction of the quantum fields. We find that, by entangling two CFTs, each one living in a separate time machine spacetime, we can construct a state that is dually described by a traversable wormhole in the bulk. So, apparently one could use quantum entanglement to teleport an excitation to the CTC region \emph{of another time machine}: instead of crossing a divergent stress at the boundary, one `takes a dive below' into the bulk, to emerge at the CTC region in the other boundary. However, we will argue that this mechanism only works to leading order in the $1/N^2$ (or $1/c$) expansion of the holographic CFT. The leading $1/c$ corrections, described by bulk quantum effects, will destroy the entangling link between the two time machines.

These results show new mechanisms that quantum theory deploys to avoid chronology violations, but we are aware that they cannot be the final answer. As we mentioned above, a divergent stress tensor coupled to dynamical gravity, no matter how weakly, will in the course of time significantly alter the geometry and thus  gravitational backreaction must be taken into account. With the understanding gained in this article, we will, in forthcoming work, extend the use of holography to investigate these effects \cite{Emparan:2021yon,future}.

The paper is organized as follows: we introduce in Sec.~\ref{sec:intuition} a simple, intuitive toy model for a time machine spacetime, given by the Misner-AdS geometry; in Sec.~\ref{sec:bulk}, we construct the bulk dual geometry from the AdS$_2$ boundary spacetime, and we discuss its numerous features, including the geometric way in which chronology protection is implemented; we continue with the analysis of the bulk geometry in Sec.~\ref{sec:patches}, and we classify the different states of the boundary CFT;  in Sec.~\ref{bulk of flat misner}, we obtain the dual geometry for a flat Misner spacetime; Sec.~\ref{sec:zerostress} discusses a special class of states with zero stress-energy tensor, and their relation to similar states found in free field theory; 
in  Sec.~\ref{sec:wormhole} we discuss how the same AdS$_2$ time machine  spacetime naturally emerges in wormhole setups, and how our results affect the physics of such traversable wormholes; we finish in Sec.~\ref{sec:final} and we discuss future directions; in App.~\ref{app:B}, we present embeddings of the geometries that we study; in App.~\ref{app:C}, we review the bulk reconstruction procedure for AdS$_3$ and apply it to our problem; App.~\ref{app:Pmap} presents a map of one of our bulk geometries to Poincar\'e-AdS$_3$, which complements the ones given in the main text.

\section{Building intuition: Misner spacetimes}
\label{sec:intuition}

We will now discuss a simple model of a time machine that shows many of its features in a very clear and appealing manner. It admits straightforward generalizations to higher dimensions \cite{Li_1999,Emparan:2021yon}.

\subsection{Constant time shift: the tilted cylinder}

As a preliminary, consider a `tilted cylinder' spacetime. The metric
\beq\label{tiltcyl}
ds^2 = -(d\tau+v d\psi)^2+d\psi^2\,,
\eeq
with constant $v$ is locally flat space, as one can see by changing
\beq
\tau=t-v\psi
\,.
\eeq
If we identify
\beq
\psi\sim\psi +L
\eeq
at constant $\tau$, this amounts to identifying 
\beq\label{timeshift}
\psi\sim\psi +L\,,\qquad t\sim t+v L\,.
\eeq
We can think of the spatial circle as the loop threading a wormhole, and then \eqref{timeshift} indicates that the mouths are connected with a constant time delay $\Delta t=v L$. It can also be seen as the result of making identifications along the circle in a frame where the circle is moving with velocity $v$. 

As long as $0\leq v<1$ the identification is along a spacelike direction, and the spacetime \eqref{tiltcyl} is not a time machine.

\subsection{Growing time shift: Misner-AdS spacetime}\label{subsec:misnerads}

\begin{figure}[t]
        \begin{center}
        \begin{subfigure}{0.5\linewidth}
        \centering
         \includegraphics[width=0.8\textwidth]{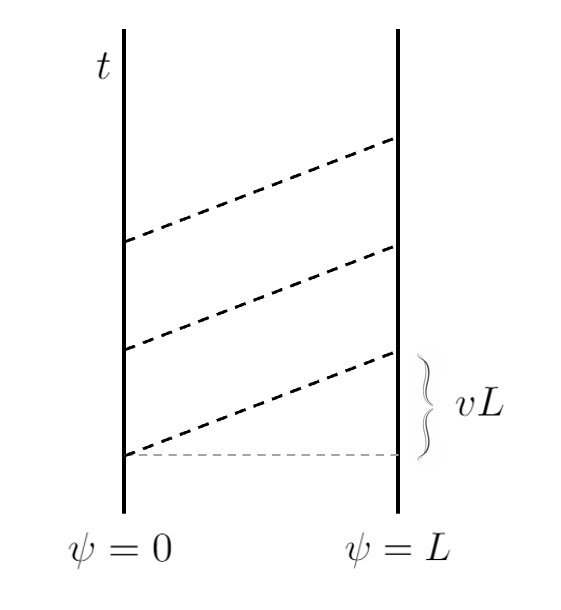}\quad
                \end{subfigure}
        \begin{subfigure}{0.42\linewidth}
        \centering
         \includegraphics[width=0.8\textwidth]{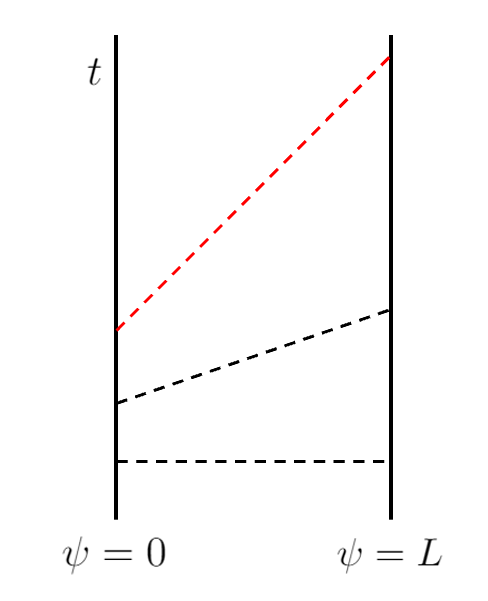}\quad
                \end{subfigure}
        \end{center}
        \caption{\small {\bf{Left}}: Tilted cylinder with $v = const$. The sides $\psi=0$ and $\psi=L$ are identified along the dashed lines, at the same time $\tau$ but different time $t$. {\bf{Right}}: Misner-AdS$_2$ cylinder with linearly increasing $v = a\tau$. The red line indicates that lightlike-separated points are identified, and therefore a closed null curve is present. At later times, closed timelike curves appear.}
        \label{fig:1}
\end{figure}

Now we build a cylinder with a tilt $v$ that grows linearly in time, with acceleration $a$ (see Fig.~\ref{fig:1}) so that
\beq\label{misnerads2}
ds^2 = -(d\tau+ a \tau\, d\psi)^2+d\psi^2\,.
\eeq

When we do this, the geometry is no longer flat space, but instead AdS$_2$ with radius $1/a$. This can be seen by changing
\beq
t= \tau\, e^{a\psi}\,,\qquad x=\frac1{a} e^{a\psi}\,,
\eeq
i.e.,
\beq
\tau=\frac{t}{a x}\,,\qquad \psi =\frac1{a}\ln ax\,,
\eeq
so the metric becomes AdS$_2$ in Poincar\'e coordinates,
\beq\label{pads2}
ds^2=\frac{-dt^2+dx^2}{a^2 x^2}\,.
\eeq
Notice that $\psi\in (-\infty,+\infty)$ covers the entire Poincar\'e patch $x\in (0,\infty)$. But we are interested in making $\psi$ into a circle, identifying it periodically along lines of constant $\tau$, i.e.,
\beq\label{psicircle}
(\tau,\psi)\sim (\tau,\psi +L)\,.
\eeq
With this identification, the geometry \eqref{misnerads2} is known as the Misner-AdS$_2$ spacetime. Writing the metric as
\beq
ds^2=-d\tau^2-2a \tau\, d\psi\, d\tau +(1-a^2 \tau^2)d\psi^2
\eeq
makes clear that $\partial/\partial\psi$ becomes timelike, and hence CTCs appear, when
\beq
|\tau|>\frac1{a}\,.
\eeq
There are two regions with CTCs, one in the future and another in the past, separated from the non-CTC region by respective chronological horizons $\tau=\pm 1/a$.  Our intention is to view this as model for a spacetime that starts at $\tau=0$ without any CTCs, and then evolves a growing time shift until it becomes a time machine. We will not be much interested in the time machine in the past, but our results readily extend to it.

\begin{figure}[t]
        \begin{center}
        \begin{subfigure}{0.52\linewidth}
        \centering
         \includegraphics[width=0.8\textwidth]{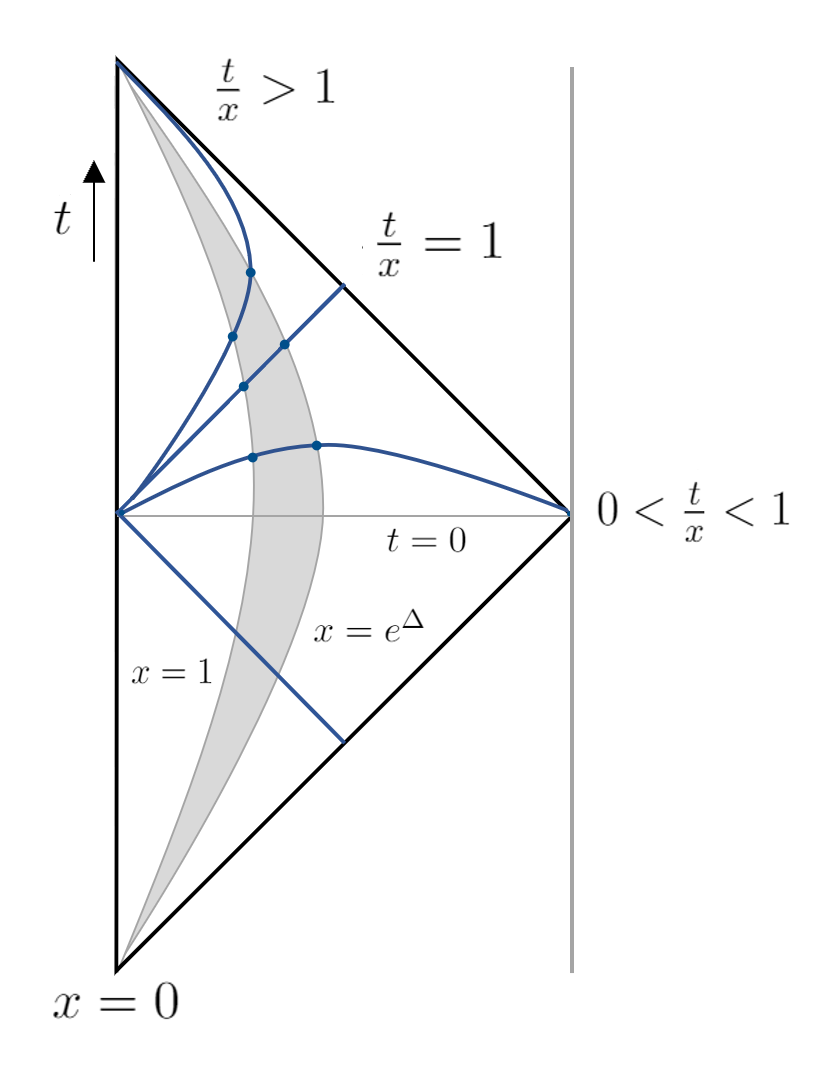}\quad
                \end{subfigure}
        \begin{subfigure}{0.44\linewidth}
        \centering
         \includegraphics[width=0.8\textwidth]{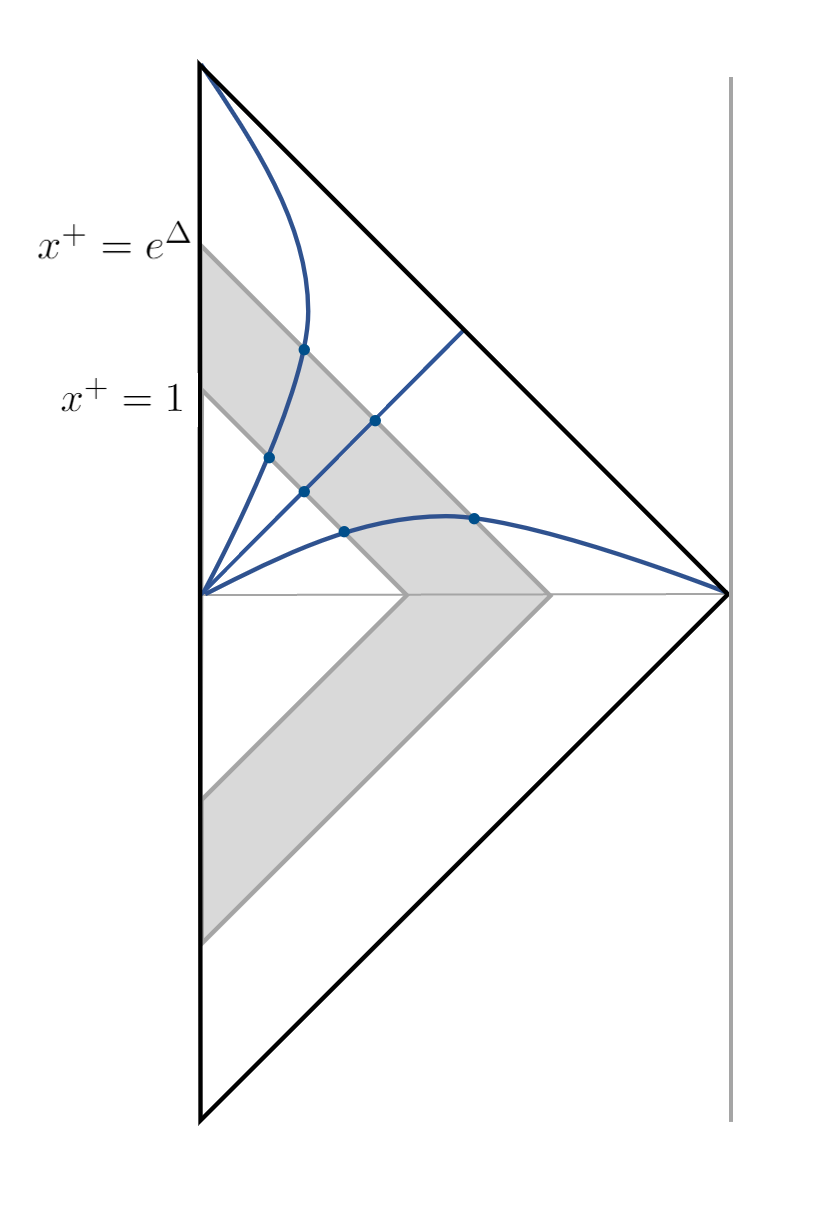}\quad
                \end{subfigure}
        \end{center}
        \caption{ \small Different choices for the fundamental region of the Misner-AdS$_2$ spacetime (shaded in gray) when embedded in the Poincar\'e-AdS$_2$ half-diamond, \eqref{pads2} or \eqref{same}. Blue dots indicate identified points, compactifying the gray strip into a cylinder. The identifications are along a light ray at the chronological horizon $t/x=1$, along timelike curves in the upper wedge $t/x>1$, and along spacelike slices in $t/x<1$.
        {\bf{Left}}: The fundamental region is given in terms of $(t,x)$ coordinates, as in \eqref{fundx}. \small {\bf{Right}}: The fundamental region is defined through $x^\pm$ coordinates, as in \eqref{reg2}.}
        \label{fig:2}
\end{figure}

We should also note that \eqref{pads2}, with the identification of points
\beq\label{idpads2}
(t,x)\sim e^\Delta(t,x)
\eeq
implied by \eqref{psicircle}, 
is exactly the geometry that one finds in a wormhole time machine model spacetime where time runs at different rates near each of the two mouths, as we will review in Sec.~\ref{sec:wormhole}. The fundamental region can be chosen in many ways, but a convenient one is (see Fig.~\ref{fig:2})
\beq\label{fundx}
x\in (1,e^{\Delta})\,,\qquad  t\in (-\infty,\infty)\,,
\eeq
and at the region boundaries the coordinates jump as in
\beq
(t,1)\sim (e^\Delta t,e^\Delta)\,. \label{reg1}
\eeq

Up to an overall scale, the only parameter of the geometry is the dimensionless number
\beq
\Delta = a L\,.
\eeq
It measures the increase rate of the tilting (or redshifting rate) in units of the circle length.  For fixed circle length, having small acceleration, i.e., slow redshifting, is equivalent to small $\Delta$. Instead, in the following we will fix the AdS radius to one, i.e.,
\beq
a=1\,,\qquad L=\Delta\,,
\eeq
so that reducing $\Delta$ shrinks the circle length. However, we will still refer to $\Delta\ll 1$ as the \textit{slow time machine} limit.

A convenient representation is obtained working in the null coordinates $x^\pm$, given by
\begin{equation}\label{nullads}
    x^- = t - x, \hspace{15pt} x^+ = t + x,
\end{equation}
so that
\begin{equation}
    ds^2 = -\frac{4}{(x^+ - x^-)^2} dx^+ dx^-\,, \label{same}
\end{equation}
and the identifications are
\beq
x^\pm \sim e^\Delta x^\pm\,.
\eeq
A possible fundamental region (different than \eqref{fundx}) is\footnote{This is for positive times. For negative ones, we take $x^-\in (-e^\Delta,-1)$.}
\beq
x^+\in (1,e^\Delta)\,,\qquad x^-< x^+\,, \label{reg2}
\eeq
(the latter is equivalent to $x>0$) with boundary conditions
\beq
(1,x^-)\sim (e^\Delta,e^\Delta x^-)\,.
\eeq

The geometry has a Killing vector
\beq\label{killingk}
k=\frac{\pd}{\pd \psi}=t\frac{\pd}{\pd t}+x\frac{\pd}{\pd x}=x^+\frac{\pd}{\pd x^+}+x^-\frac{\pd}{\pd x^-}\,,
\eeq
whose orbits are periodic under the identifications we are making. These same identifications break other symmetries of AdS$_2$, in particular the time translation symmetry generated by $\partial/\partial t$. As a result, the geometry is time-dependent in a manner made manifest by the growing tilt of the Misner-AdS cylinder. 

\begin{figure}[t]
    \centering
    \includegraphics[width=0.45\textwidth]{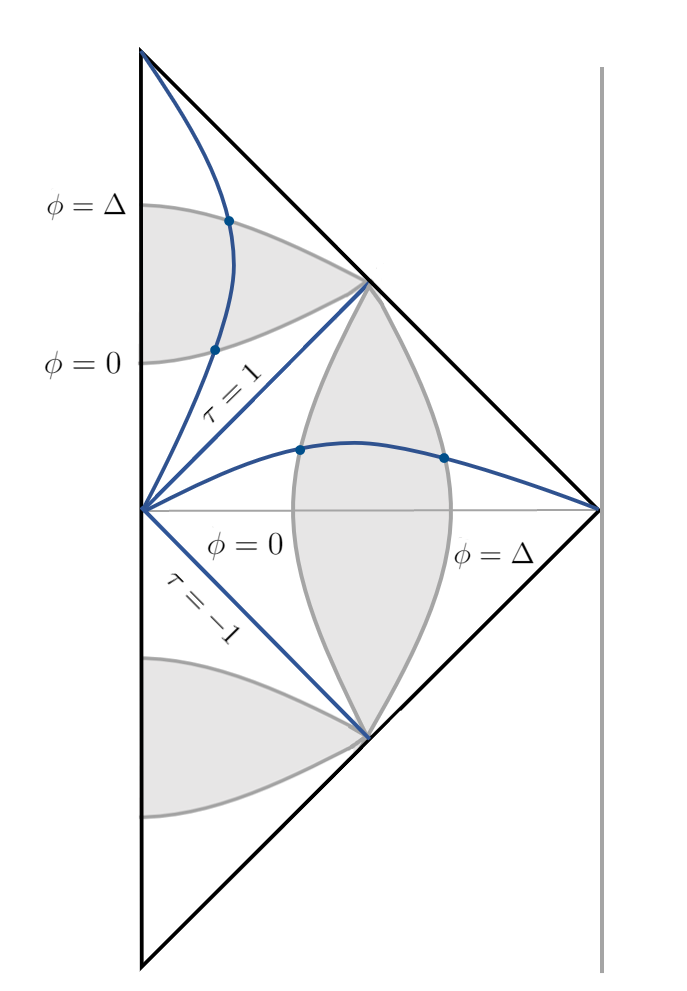}
    \caption{\small The fundamental regions for the lower and upper patches in $(\tau,\phi)$ coordinates, both shaded in gray. The blue dots correspond to identified points. \textbf{Lower patch}: Determined by $\tau^2<1$, $\phi \in [0,\Delta]$. Blue line is constant $\tau <1$ spatial slice. \small \textbf{Upper patch}: Described with $\tau>1$, and with $\phi \in [0,\Delta]$ now as compact time coordinate. Blue line indicates constant $\tau >1$ timelike trajectory. The chronological horizons are reached for $\tau=\pm 1$.}
    \label{fig:3}
\end{figure}

Finally, let us present another set of coordinates in which the metric becomes diagonal but in a different form than \eqref{pads2} (see figure \ref{fig:3}). Namely, keep $\tau$ but change $\psi\to\phi$ so that 
\beq
d\psi=d\phi +\frac{\tau}{1-\tau^2}d\tau\,,
\eeq
i.e.,
\beq\label{psiphi}
\psi = \phi -\ln\sqrt{|1-\tau^2|}\,.
\eeq
The metric of Misner-AdS$_2$ now becomes
\beq\label{tauphimetric}
ds^2=-(\tau^2-1)d\phi^2+\frac{d\tau^2}{\tau^2-1}
\eeq
with periodicity $\phi\sim \phi+\Delta$ along the orbits generated by the Killing vector
\beq
k=\frac{\partial}{\partial\phi}\,.
\eeq
The form of the metric \eqref{tauphimetric} is like Rindler-AdS$_2$, but in the latter, conventionally one considers the region $\tau^2>1$, so that $\tau$ is a spatial coordinate and $\phi$ is the time. Therefore the time machine part of Misner-AdS$_2$ is Rindler-AdS$_2$ with closed timelike orbits, and the Rindler acceleration horizons $\tau=\pm 1$ correspond to the chronology horizons of the Misner space. In the CTC region, the spatial sections are non-compact.
The non-CTC region $\tau^2<1$ is time-dependent with circular spatial sections. It is a Milne-AdS$_2$ cosmology\footnote{The Milne universe is a cosmology obtained as a wedge of Minkowski space that either expands or collapses. Milne-AdS$_2$ is a bouncing universe that first expands and then collapses.} that is spatially compact but is free of CTCs. 

In these coordinates the CTC and non-CTC regions are described as separate coordinate patches. We will refer to the region $\tau>1$ as the `upper patch' and to $-1<\tau<1$ as the `lower patch'. Although there is also a past CTC region for $\tau<-1$, we will mostly ignore it in our discussion.

\subsection{Flat limit: Misner spacetime}\label{subsec:toflat}

Let us now zoom in on the geometry near the (future) chronological horizon at $\tau=1$. To this effect, we make
\beq\label{toflatmisner}
\tau=1+\epsilon T\,,
\eeq
where $T$ is a new coordinate, and rescale the metric such that in the limit $\epsilon\to 0$, \eqref{misnerads2} becomes
\beq\label{flatmisner}
\epsilon^{-1} ds^2\to -2 dT d\psi-2Td\psi^2\,.
\eeq
When $\psi$ is a periodic coordinate, this is the geometry of Misner space as originally introduced in \cite{Misner:1965zz}. It is a discrete quotient of Minkowski space that contains CTCs when $T>0$. This structure is inherited from the original Misner-AdS spacetime. We can also obtain the limit in the null coordinates \eqref{nullads} by taking
\beq\label{toflatnull}
x^+ =-\frac2{X^+}\,,\qquad x^-=\frac{\epsilon}{2} X^-\,,
\eeq
such that \eqref{same} becomes
\beq
\epsilon^{-1}ds^2\to -dX^+ dX^-\,.
\eeq
This is made into a time machine when we identify along Minkowski boosts,
\beq\label{boostident}
X^+\sim e^{-\Delta}X^+\,,\qquad X^-\sim e^\Delta X^-\,.
\eeq
The geometry is contained in the half-space $X^+<0$, and the non-CTC and CTC regions correspond to $X^-<0$ and $X^->0$, respectively, with chronological horizon at $X^-=0$. In the coordinate $\phi$ of \eqref{psiphi}, now defined as
\begin{equation}
    \phi=\psi+\frac12\ln |T|\,,
\end{equation}
the metric is
\begin{equation}
    ds^2=\frac{dT^2}{2T}-2Td\phi^2\,,
\end{equation}
with CTCs when $T>0$.

The identification \eqref{boostident} along boosts allows to present Misner space as a model of the time machine that results when the two mouths of a wormhole approach and pass by each other with relative velocity $\tanh\Delta$. Although flat Misner space is slightly simpler than Misner-AdS, the latter is of interest to us as a more general class of spacetimes that evolve CTCs, and the model of a broad family of wormhole time machines, as we argue in Sec.~\ref{sec:wormhole}.

\subsection{Motion in Misner spacetimes}\label{sec:motion}

The metrics \eqref{misnerads2} and \eqref{flatmisner}, in coordinates $(\tau,\psi)$, extend analytically the manifold across the chronology horizon, but nevertheless there are future-incomplete null geodesics. These features can be easily inferred from the Penrose diagrams in Fig.~\ref{fig:1}, but let us see them explicitly in the Misner coordinates $(\tau,\psi)$. The `right-moving' null geodesics are
\begin{equation}
    \tau=1-c\lambda\,,\qquad \psi=-\ln(-\lambda)
\end{equation}
with affine parameter $\lambda\in(-\infty,0)$ and constant $c\in\R$. The future chronology horizon $\tau=+1$ is reached when $\lambda\to 0$, after circling an infinite number of times around $\psi$. For $c<0$ one reaches $\tau=1^-$ advancing towards the future, and for $c>0$ one reaches $\tau=1^+$ moving backwards in time. In both cases the geodesics terminate at the future chronology horizon at $\lambda=0$, so they are incomplete. Instead, the left-moving null geodesics are
\begin{equation}
    \tau=-1+c \lambda\,,\qquad \psi=-\ln \lambda\,,
\end{equation}
with $\lambda\in(0,\infty)$. When $c>0$ they smoothly cross the horizon $\tau=1$ to the future at finite affine parameter $\lambda=2/c$.\footnote{They also terminate at the past chronology horizon, but we are not interested in this.}

Particle trajectories are also easy to analyze. The time machine region can be accessed, since there exist timelike geodesic segments that cross the future horizon in finite proper time. There are also complete non-geodesic particle trajectories, e.g. at constant $x$, that cross both the past and future horizons.
Thus, although Misner spacetimes are sometimes regarded as singular, they are better referred to as quasi-regular. 

A particle in geodesic motion in the circle, even if it starts at rest, acquires an increasing non-zero velocity. The particle gains more and more energy by `falling' inside the circle due to its tilting. In the Poincar\'e representation \eqref{pads2}, when the particle falls to $x=e^{\Delta}$, the identifications \eqref{reg1} lift it up in the gravitational field to $x=1$. In the flat Misner spacetime, the energy gain is viewed as due to Lorentz boosting. For each turn around the circle, the energy of the particle increases by a factor $e^{\Delta}$. When the time machine is about to form, i.e., the future chronological horizon is approached, the energy gain diverges. When a quantum CFT is put in this spacetime, its fluctuations gain energy in the same manner. This is natural, since the field is excited by the time dependence of the geometry.

\section{Holographic dual of the time machine}
\label{sec:bulk}

After characterizing the geometry of the time machine  spacetime, we now turn to the evolution of interacting quantum fields in it. We employ the holographic approach in which a strongly coupled conformal field theory with large central charge $c$ is described through a dual three-dimensional bulk geometry with the Misner-AdS$_2$ geometry at its boundary.

\subsection{Stress tensor}

In order to reconstruct this bulk, we first constrain the form of the quantum stress tensor. For a CFT with central charge $c$, it can be decomposed into a traceless component and an anomalous trace term
\beq
\langle T_{ij}\rangle =\langle \hat{T}_{ij}\rangle +\frac{c}{48\pi}R g_{ij}
\eeq
with
\beq
\langle \hat{T}^i{}_i\rangle=0\,,\qquad R=-2\,,
\eeq
the latter being the scalar curvature of unit-radius AdS$_2$. Our goal is to obtain the traceless part $\langle \hat{T}_{ij}\rangle$.

Conformal invariance, together with the time-reversal symmetry of the background geometry, impose that the stress tensor, in the null coordinates of \eqref{same}, takes the form
\beq\label{stressalpha}
\langle \hat{T}_{ij}\rangle dx^i dx^j=-\alpha \frac{c}{24\pi}\lp \lp \frac{dx^+}{x^+}\rp^2+\lp \frac{dx^-}{x^-}\rp^2\rp\,,
\eeq
with a coefficient $\alpha$ that, in principle, can be a function of the coordinate ratio $x^+/x^-=(\tau+1)/(\tau-1)$. Indeed, the translation symmetry along $\psi$ leaves only the possibility that it is a function of $\tau$. Then the covariant conservation condition $\nabla_i \langle \hat{T}^{ij}\rangle=0$ requires that $\alpha$ be constant.

In $(\tau,\phi)$ coordinates, the stress tensor is diagonal,
\beq\label{stressalpha2}
\langle \hat{T}^{\tau}_\tau\rangle =-\langle \hat{T}^{\phi}_\phi\rangle =\frac{c}{12\pi}\frac{\alpha}{ 1-\tau^2}\,,
\eeq
and therefore does not have any energy-flow component. 
Thus, these coordinates are adapted to a frame comoving with the CFT. In this frame, when $\tau^2<1$ and $\phi$ is a spatial direction, the stress tensor is time-dependent but spatially homogeneous. When $\tau^2>1$ and $\phi$ is timelike and $\tau$ spacelike, the stress-tensor is static but inhomogeneous. In both regions, the stress tensor diverges at the chronology horizons.

The nature of this quantum stress tensor changes depending on the value of $\alpha$. In the non-CTC region, where the system is more clearly understood, a state of the CFT for which $\alpha>0$ has negative energy density and positive tension, which we interpret as being dominated by the Casimir stress-energy of the fields in a time-dependent circle (the closed Milne spacetime). Instead, in states with $\alpha<0$ there is positive energy density and pressure, which are more appropriately viewed as of thermal origin. Below we will elaborate further on these interpretations and see that, when $\alpha<1/2$, the two components, Casimir and thermal, are present, changing their relative dominance when $\alpha$ flips sign.

Our strategy now is to employ the standard bulk reconstruction procedure \cite{de_Haro_2001} in order to obtain an exact solution of the three-dimensional Einstein-AdS equations that has the Misner-AdS$_2$ geometry at its boundary, and \eqref{stressalpha} as its holographic stress tensor. 

\subsection{Bulk metric and regularity}

In App.~\ref{app:C}, we explain the procedure of bulk reconstruction in AdS$_3$/CFT$_2$, and apply it to the Misner-AdS$_2$ time machine spacetime with stress tensor \eqref{stressalpha}. In the null coordinates of \eqref{same}, the bulk metric that we obtain takes the form
\begin{align}
ds^2
=\frac{\ell^2}{z^2}\Biggl[dz^2 & - \lp \lp 1+\frac{z^2}{4}\rp^2\frac{4}{(x^+-x^-)^2} +\frac{\alpha^2}{16}\lp\frac1{x^+}-\frac1{x^-}\rp^2 z^4\rp dx^+ dx^-\nn\\
&-\alpha\frac{z^2}{2}\lp 1+\frac{z^2}{4}\rp\lp \lp \frac{dx^+}{x^+}\rp^2+ \lp  \frac{dx^-}{x^-}\rp^2\rp\Biggr]\,.\label{AdStimemach}
\end{align}
The coordinate $z$ extends from the boundary at $z=0$ into the bulk with $z>0$, in a manner that we will explore in detail below. Some expressions become simpler in terms of the proper distance $\sigma=\ln(2/z)$, and we will use it occasionally.

We can also transform \eqref{AdStimemach} into any of the coordinates for the boundary metric  introduced in the previous section. In the coordinates $(\tau,\phi)$ it becomes a diagonal metric,
\begin{align}
ds^2 &= \frac{\ell^2}{z^2}\Biggl[ dz^2 -\lp 1+\frac{z^2}{4}+\frac{\alpha}2 \frac{z^2}{1-\tau^2}\rp^2 \frac{d\tau^2}{1-\tau^2}+\lp 1+\frac{z^2}{4}- \frac{\alpha}2\frac{z^2}{1-\tau^2}\rp^2 (1-\tau^2)d\phi^2\Biggr]\,.\label{tauphibulk}
\end{align}
Recall that $\tau$ is timelike and $\phi$ spacelike when $\tau^2<1$, and they reverse roles when $\tau^2>1$. It is easy to see that curves of constant $\tau$ and of constant $\phi$ are geodesic.

This geometry is, by construction, locally AdS$_3$ for any value of $\alpha$, but global bulk regularity is not guaranteed. To begin the interpretation of the geometry, note that the boundary metric and the CFT stress tensor are independent of $\phi$, and therefore
\beq
k=\frac{\pd}{\pd\phi}
\eeq 
is also a Killing vector of the bulk. Its norm is given by
\beq 
|k|^2=g_{\phi\phi}=\frac{\ell^2}{z^2}\frac1{1-\tau^2}\lp (1-\tau^2)\lp 1+\frac{z^2}{4}\rp-\frac{\alpha}2 z^2\rp^2\,,
\eeq
which implies that $k$ is spacelike for $|\tau|<1$ and timelike for $|\tau|>1$. The bulk then contains CTCs in precisely the same range of $\tau$ as in the boundary. However, when $|\tau|=1$ the vector $k$ is null only when $z=0$. At any finite distance $z$ inside the bulk, $k$ diverges when $|\tau|=1$. Therefore, the chronology horizon is only present at the boundary.

There are, however, points in the bulk where $|k|=0$ while $k$  is not null, but it becomes the zero vector. This happens where
\beq\label{zmaxt}
z=z_\textrm{max}=2 \sqrt{\frac{1-\tau^2}{2\alpha-1+\tau^2}}\,.
\eeq
When $k$ is spacelike, i.e., $|\tau|<1$, $z_\textrm{max}$ is real when
\beq\label{alpharange}
\alpha>\frac12\,.
\eeq
Then, in the region $0<z<z_\textrm{max}$ the vector $k$ reaches a fixed point-set at $z=z_\textrm{max}$. The absence of conical singularities at these points demands that we fix the periodicity $\Delta$ of the orbits of $k$ to the value
\begin{align}\label{Deltaalpha}
\Delta=\lim_{|k|\to 0} \frac{2\pi} {\lp \partial_\mu |k|\partial^\mu |k|\rp^{1/2}}=\frac{2\pi}{\sqrt{2\alpha-1}}\,,
\end{align}
or, equivalently,
\beq\label{alphaDelta}
\alpha=\frac12+\frac{2\pi^2}{\Delta^2}\,.
\eeq
The condition \eqref{alpharange} then follows whenever $\Delta\neq 0$. Observe that, since $\alpha>0$, these states have what we described as negative Casimir energies. This is what we expect in  the physical situation we are mostly interested in, namely, starting at $\tau=0$ in the ground state of the CFT in a circle, with negative Casimir energy, and then evolving in time towards the formation of a time machine.
Values of $\alpha$ outside the range \eqref{alpharange} can also yield sensible bulks, but we will see that they must be interpreted differently. 

For the situation of our main interest, this completes the bulk construction, at least as long as we are only concerned with the regime before CTCs are developed. We have obtained that stress tensor in the ground state of the strongly coupled CFT in the Misner-AdS$_2$ geometry, with a given shifting rate $\Delta$, and for times $|\tau|<1$, has traceless component 
\beq\label{TDelta}
\langle \hat T_{ij}\rangle dx^i dx^j=-c\lp \frac1{48\pi}+\frac{\pi}{12\Delta^2}\rp \lp \lp \frac{dx^+}{x^+}\rp^2+ \lp  \frac{dx^-}{x^-}\rp^2\rp\,,
\eeq
where the central charge is $c=3\ell/2G$.

 \paragraph{``Slow time machine'' regime.}


It is illustrative to examine in more detail the initial, small redshifting regime. For this purpose, we focus on times shortly after $\tau=0$, rescale $\psi$ to keep the circle length finite, and also rescale $z$, whose range shrinks for small $\Delta$ (i.e., large $\alpha$). All of this is achieved by changing
\beq
(\tau,\psi,z)\to \frac{\Delta}{2\pi}(\tau,\psi,z)\,,
\eeq
and then expanding for small $\Delta$. Observe that the new $\psi$ has periodicity $2\pi$. If we furthermore introduce, for mere convenience, the proper bulk distance coordinate $\sigma=\ln(2/z)$, then the bulk geometry becomes
\begin{align}
ds^2=& \,d\sigma^2-\cosh^2\sigma\, d\tau^2+\sinh^2\sigma\, d\psi^2
-\frac{\Delta}{\pi}\tau\sinh^2\sigma\, d\tau d\psi+O(\Delta^2)\,,\label{slowtm}
\end{align}
for which the holographic stress tensor is
\begin{align}\label{Tslow}
\langle T_{ij}\rangle dx^i dx^j
&=-\frac{c}{24\pi}\lp d\tau^2+d\psi^2 -\frac{2\Delta}{\pi} \tau\, d\tau d\psi +O(\Delta^2)\rp
\end{align}
 (the anomalous trace vanishes to linear order in $\Delta$). 
The leading order geometry in \eqref{slowtm} is the global AdS$_3$ spacetime, which is the bulk dual of the CFT in a circle with Casimir energy $\langle T_{tt}\rangle=\langle T_{\psi\psi}\rangle<0$. Thus we recover the result we anticipated for the initial state of the CFT. The correction to the metric proportional to $\Delta$ gives the bulk dual of the energy flux $\langle T_{t\psi}\rangle$ that is created by the growing tilt in the Misner-AdS$_2$ cylinder. 

In the interpretation of Sec.~\ref{sec:wormhole}, the geometry \eqref{slowtm} captures holographically the quantum effects induced when beginning to create a time differential between the two mouths of a wormhole.

\paragraph{Comparison with free CFT.}

The dependence of the stress tensor on $\Delta$ differs, as would be expected, from the one obtained in \cite{PhysRevD.43.3878} for a free conformal scalar field, given by \eqref{stressalpha} with $c=1$ and 
\begin{equation}
    \alpha_f =\sum_{n=1}^\infty \frac{3}{\sinh^2(n\Delta/2)}. \label{fro}
\end{equation}
The infinite sum appears because of the method of images used to solve the field theory in a discrete quotient of AdS$_2$. The analysis of $\alpha_f(\Delta)$ is simpler if we rewrite the sum in terms of q-trigamma functions, but one can already see some of its distinct features in its current form. First, let us note that for small values of $\Delta$,
\begin{equation}
    \alpha_f \simeq \frac{12}{\Delta^2}\sum_{n=1}^\infty \frac1{n^2}= \frac{2\pi^2}{\Delta}\,,
\end{equation}
which matches exactly \eqref{alphaDelta} in the same limit. This is the regime of the ``slow time machine'' analyzed above, in which the effects can be captured perturbatively in $\Delta$ in both cases. However, for large values of $\Delta$, the function $\alpha_f(\Delta$) falls off exponentially to zero; large $\Delta$ corresponds to strong time machine effects. We see that this behaviour is quite different  compared to our case, where for large $\Delta$ we obtain a constant contribution, as can be seen from \eqref{TDelta}. Therefore, the strong coupling contributes to a larger field excitation for fast time machine formation, compared to the free field analysis. 

Next we turn to a more significant effect of the strong coupling of the holographic quantum fields.

 \paragraph{Chronology self-protection.}
    
Observe that \eqref{zmaxt} implies that as $\tau$ increases from $\tau=0$ towards the chronological horizon at $\tau=1$, the value of $z_\textrm{max}$ decreases: The extent of the bulk away from the boundary shrinks, until the moment $\tau=1$ when the bulk size vanishes. Then, continuing to $\tau>1$, the bulk reappears, and indeed there is no maximum value of $z$ anymore, since $z$ can run from $0$ to $\infty$. Below we will further clarify what is going on in this peculiar geometry, but for now let us emphasize one important consequence of the closing up of the bulk at $|\tau|=1$: 
\begin{quote}
It is impossible to cross from the chronologically normal region to the time machine region following a trajectory that remains at finite spacetime distance inside the bulk.
\end{quote}
Or, in dual terms:
\begin{quote}
An excitation of the CFT cannot cross the chronological horizon.
\end{quote}

We emphasize, as we discussed in the introduction, that this chronology protection is invisible to free fields in the fixed background geometry of Misner-AdS$_2$. But in our setup, if we consider a wavepacket of the conformal fields, its interaction with the background state of the field is holographically modelled as propagation in the bulk geometry. The disconnection between the two sides of the bulk represents holographically that the divergence of the fields at the chronology horizon impedes the passage to the other side of any field excitations---indeed, of any other system that interacts with the CFT affecting only weakly its state. This happens even in the absence of gravitational backreaction on the 2D geometry, which our model does not include. 

The holographic quantum fields are strictly infinitely strongly coupled, but it is possible that finite self-interactions of the fields lead to the same conclusion. This would be very hard to see with conventional techniques. Perturbative self-interactions may hint at this, but their study will be more involved and necessarily less controlled.

Holography then suggests that the role of strong gravitational effects for enforcing chronology protection may be less important than is often assumed. Our detailed analysis of the holographic bulks in the next section will emphasize this conclusion in other ways.

\section{Patching up the bulk geometry}
\label{sec:patches}

Let us continue the analysis of the bulk metrics \eqref{tauphibulk}.
For any value of $\alpha$, they are locally equivalent to AdS$_3$, and therefore it must be possible to find a change of coordinates that transforms them into other more familiar forms of AdS$_3$. 

For simplicity we will set the AdS radius to $\ell=1$. We leave the coordinate $\phi$ unchanged and transform the coordinates $(z,\tau)$ into new ones $(r,t)$, defined by
\begin{align}
r^2&= \frac{|1-\tau^2|}{z^2}\lp 1+\frac{z^2}{4}- \frac{\alpha}2\frac{z^2}{1-\tau^2}\rp^2
\,,\label{rtauz}\\
t&=\frac1{2}\ln\left|\frac{1+\tau}{1-\tau} 
\lp \frac{4(1-\tau^2)-z^2\lp \sqrt{1-2\alpha}-\tau\rp^2}{4(1-\tau^2)-z^2\lp \sqrt{1-2\alpha}+\tau\rp^2}\rp^{\frac1{\sqrt{1-2\alpha}}}
\right|\,,
\end{align}
which take our bulk metric \eqref{tauphibulk} into
\begin{equation}\label{genads}
    ds^2=\kappa\lp (r^2+\kappa(1-2\alpha)) dt^2- r^2 d\phi^2\rp +\frac{dr^2}{r^2+\kappa(1-2\alpha)}\,,
\end{equation}
where we have introduced a parameter
\begin{equation}
    \kappa=\rm{sign}(\tau^2-1)=
    \begin{cases}
    +1\quad &\text{upper patch (CTC region)\,,}\\
    -1\quad &\text{lower patch (no-CTC region)\,.}
    \end{cases}
\end{equation}
Bear in mind that, despite the notation, we are not preconditioning whether $t$ and $\phi$ are timelike or spacelike bulk coordinates, nor are we necessarily restricting to $r^2\geq 0$. What is common to all patches is that $\phi$ is periodically identified, $\phi\sim\phi +\Delta$, while $t$ is never periodic\footnote{Varying $\tau$ and $z$ over their allowed ranges, $t$ always runs between $-\infty$ and $+\infty$. If needed, $t<0$ can always be obtained by time reversal symmetry.}. This map is valid for all $\alpha$ without assuming that it is related to $\Delta$. When $\alpha>1/2$, we can replace $\sqrt{1-2\alpha}=i\sqrt{2\alpha-1}$, and the function $t(\tau,z)$ can be made manifestly real by rewriting it in terms of $\arctan$ functions.

We proceed by analyzing the different ranges of $\alpha$ separately.

\subsection{$\alpha>1/2$: Split-bulk chronology protection}

For the sake of clarity, we introduce
\begin{equation}
    r_0^2=2\alpha-1>0\,.
\end{equation}

\paragraph{Lower patch.} The metric \eqref{genads} for $\kappa=-1$ is
\beq\label{gads3}
ds^2=-(r^2+r_0^2)dt^2+r^2 d\phi^2+\frac{dr^2}{r^2+r_0^2}\,.
\eeq
This is the geometry of global AdS$_3$, with $-\infty<t<\infty$. The origin $r=0$ is the fixed point set of $k=\partial_\phi$, which we already identified in \eqref{zmaxt}, and regularity requires that $r_0=2\pi/\Delta$, which is the same as \eqref{alphaDelta}.

Thus we reproduce our conclusion that the state of the CFT---i.e., the holographic bulk dual with a specific value of $\alpha$ for the holographic stress tensor---in the non-CTC region of Misner-AdS$_2$ with parameter $\Delta$ is uniquely fixed. The bulk is geodesically complete, so excitations of the CFT cannot escape the region $\tau^2<1$. The chronology horizon of the boundary, $\tau^2=1$, is at $t=\pm\infty$ and therefore is never reached at any finite time in the bulk. An excitation that follows a null or timelike geodesic in the bulk will wander for an arbitrarily long proper time, or affine parameter, without ever arriving at $\tau^2=1$. From the CFT viewpoint, its energy grows unbounded as the chronology horizon is approached (since $z\to 0$), but never crosses it.

\begin{figure}
    \centering
    \includegraphics[width=.85\textwidth]{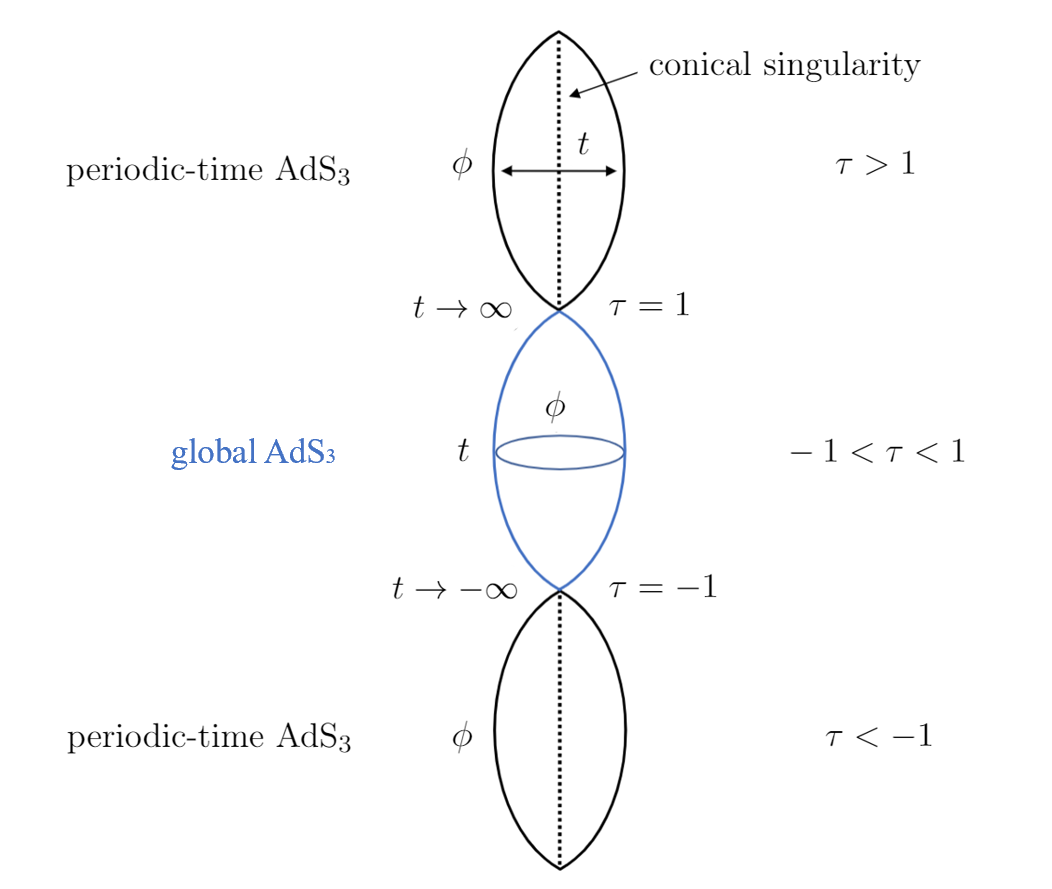}
    \caption{\small Sketch of the bulk for $\alpha>1/2$ states (not a proper Penrose diagram). The boundary is a single, complete Misner-AdS$_2$ geometry with chronology horizons at $\tau=\pm 1$, but the bulk consists of three disconnected spacetimes. For $-1<\tau<1$, it is a geodesically complete, global AdS$_3$ spacetime, with time coordinate $t\in(-\infty,+\infty)$ and without CTCs. For $|\tau|>1$, the bulk is an AdS$_3$ geometry, with CTCs since the time $\phi$ is periodically identified (not represented in the figure), and with a conical singularity at the origin $r=0$, since the `angular' coordinate $t$ runs in $(-\infty,+\infty)$.}
    \label{fig:global}
\end{figure}

It is worth noticing how by fixing a specific boundary geometry we obtain features that are not conventionally present in global AdS$_3$.
For instance, one may ask how the static bulk of global AdS$_3$ is compatible with the time-dependence of the boundary geometry. The answer is that the timelike Killing vector in the bulk does not generate time translations at the boundary, since these are broken by the asymptotic boundary conditions. The choice of a conformal factor that is singular at $\tau^2=1$  is what allows the boundary geometry to cover the two patches that are disconnected through the bulk. The coordinate $\tau$ compactifies the infinite extent of the bulk time $t$ into the finite interval $\tau^2<1$. Thus, although the lower-patch bulk is complete, its boundary geometry is not,\footnote{This is the reverse of, e.g., Poincar\'e-AdS, where the Minkowski boundary is complete but the bulk is not.} and its extension to $\tau^2>1$ involves a different bulk spacetime (Fig.~\ref{fig:global}).

Finally, observe also that global AdS$_3$ does not have any free dimensionless parameters, while our solutions do have one, namely $\Delta$. This parameter is a property of the boundary geometry; ultimately, it is introduced by the choice of conformal frame for the boundary metric.

\paragraph{Upper patch.}  With $\kappa=+1$ the metric \eqref{genads} is
\beq\label{gads32}
ds^2=(r^2-r_0^2)dt^2-r^2 d\phi^2+\frac{dr^2}{r^2-r_0^2}\,.
\eeq
If we transform 
\begin{equation}
    r^2=\tilde{r}^2+r_0^2
\end{equation}
we see that the time coordinate $\phi$ is a global time---there are no horizons---so this looks again like the metric for global AdS$_3$. However, there are two important differences. First, since $\phi$ is periodic there are CTCs, not only at the boundary but throughout the entire bulk. Second, the spatial coordinate $t$ is not periodic, since its range $(-\infty,\infty)$ is determined by that of the boundary coordinate $\tau$. Then, at $\tilde{r}=0$ there is a conical singularity (an infinite conical excess).  In this patch there is no need to fix $\alpha$ in terms of $\Delta$, but in any case the state of the CFT for $\tau^2>1$ is not well defined for any $\alpha>1/2$ due to the bulk singularity.

A curious feature is that, even if the coordinates $(t,\tilde{r},\phi)$ provide a maximal extension of the bulk, the same is not true if we use the Fefferman-Graham coordinate $z$.
It turns out that the function $\tilde{r}(\tau,z)$ takes its minimum value for
\beq\label{tohole}
z\to 0\,,\qquad \tau^2\to 1\,,\qquad \frac{z^2}{\tau^2-1}=\frac{2}{\alpha}\,.
\eeq
This minimum corresponds to $\tilde{r}=1$, which is an infinite line since $t$ is not periodic. So the region $\tilde{r}<1$ is a `hole' in the bulk in terms of the coordinate $z$. It is not clear why this happens, but Fefferman-Graham coordinates are known to not cover maximally other bulks, e.g., they fail to include the interiors of black branes. This is in any case a gauge issue, since we can always transform to $\tilde{r}$ and continue down to $\tilde{r}=0$.

\subsection{$\alpha<1/2$: Traversable bulk wormholes and entangled time machines}

For this case we define
\begin{equation}
    r_0^2=1-2\alpha>0\,.
\end{equation}

We will show that, in contrast to what we found above, now the lower and upper patches of the bulk are connected as part of a single manifold extended across the chronology horizons. However, to arrive at this conclusion we shall begin analyzing them separately.

\paragraph{Lower patch.}
Now the metric \eqref{genads} with $\kappa=-1$,
\beq\label{BTZ1}
ds^2=-(r^2-r_0^2)dt^2+r^2d\phi^2+\frac{dr^2}{r^2-r_0^2}\,,
\eeq
is that of a static BTZ black hole with horizon at $r=r_0$. Since the periodicity of $\phi$ is not $2\pi$ but $\Delta$, 
a rescaling of the coordinates must be made to obtain the canonical form of the solution, but we will not need that.


Having identified the bulk geometry as a black hole, we can now understand our previous observation that the stress tensor of states with $\alpha<0$ is of thermal type: the BTZ black hole is indeed dual to a CFT at finite temperature. So, actually, whenever $\alpha<1/2$ the CFT in Misner-AdS$_2$ has a thermal component, but when $0<\alpha<1/2$ the positive thermal energy is overwhelmed by the negative Casimir energy. There is a state with $\alpha=0$ where the two components exactly cancel each other. Below we will examine it more closely.

Usually, the BTZ black hole spacetime \eqref{BTZ1} is taken to end at $r=0$, even though the curvature remains finite. That is a chronology horizon of Misner-type, where some null geodesics become incomplete, while others cross it smoothly. In this article we are precisely interested in continuing across these surfaces. Therefore, we will extend the geometry beyond $r=0$, into a region where the bulk contains CTCs. Unsurprisingly, crossing into this region takes us to the upper patch, as we will show next.

\begin{figure}[t]
    \centering
    \includegraphics[width=0.40\textwidth]{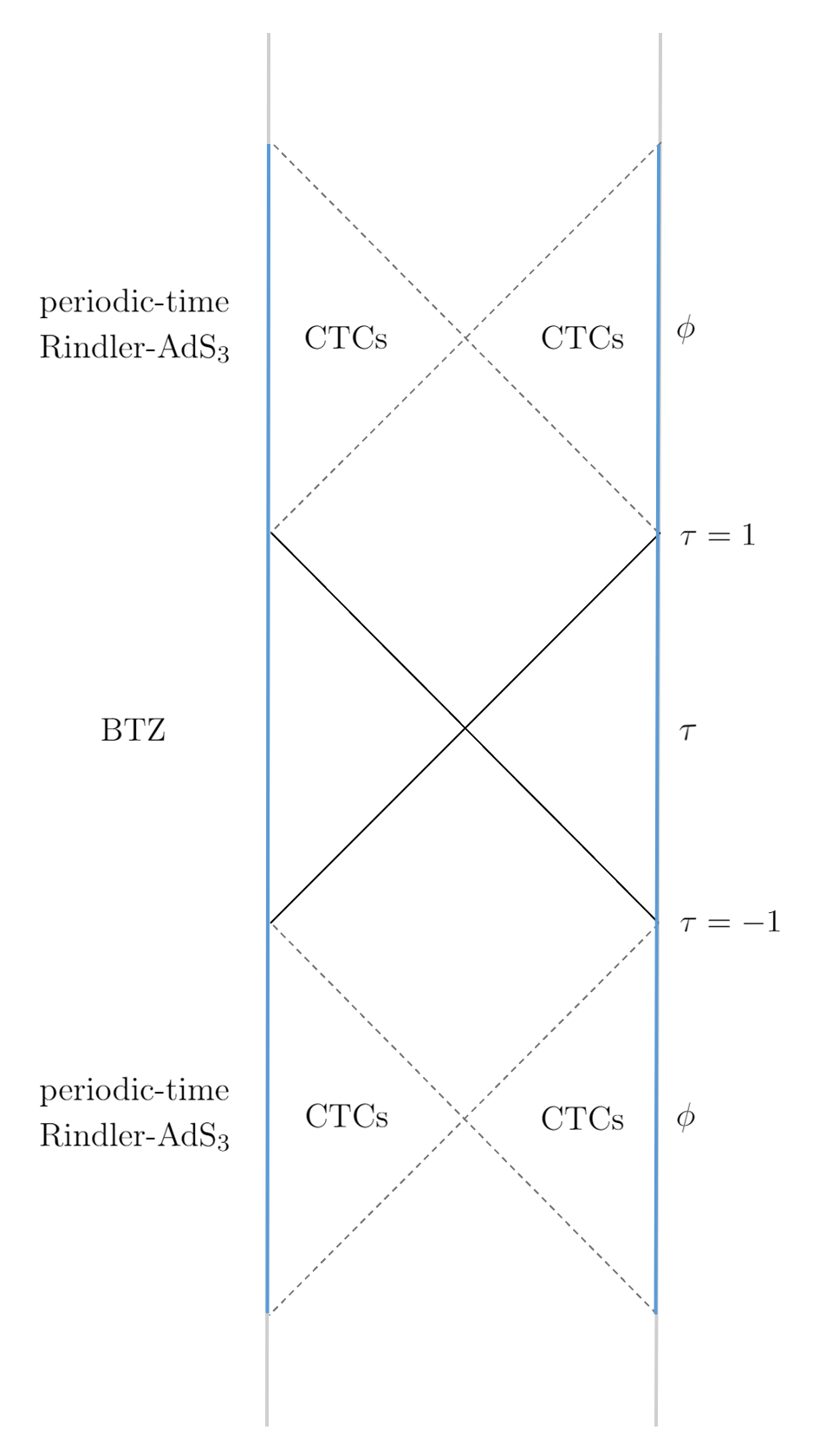}
    \caption{\small Sketch of the bulk for $\alpha<1/2$ states (not a proper Penrose diagram, since it mixes sections of constant $\phi$ for the BTZ portion, and of constant $t$ for the Rindler-AdS$_3$ portions, where the periodicity of time $\phi$ is not represented). Dashed diagonal lines represent chronology horizons, while the diagonal solid lines indicate the BTZ black hole horizons. The blue boundary lines represent two identical, complete Misner-AdS$_2$ spacetimes. Since it is possible to cross from one to the other, the bulk is a traversable wormhole. It is expected that bulk quantum effects will render the chronology horizons singular, and close up the wormhole.}
    \label{fig:penrose}
\end{figure}

\paragraph{Upper patch.} 
With $\kappa=+1$ in \eqref{genads} we get
\beq\label{Rads}
ds^2=(r^2+r_0^2)dt^2-r^2 d\phi^2+\frac{dr^2}{r^2+r_0^2}\,.
\eeq
The interpretation becomes clearer by changing
\begin{equation}
    r^2=\tilde{r}^2-r_0^2
\end{equation}
so \eqref{Rads} becomes
\beq\label{Rads2}
ds^2=\tilde{r}^2 dt^2-(\tilde{r}^2-r_0^2) d\phi^2+\frac{d\tilde{r}^2}{\tilde{r}^2-r_0^2}\,.
\eeq
This looks again like the BTZ black hole, now with time $\phi$. However, the spatial coordinate $t$ runs in $(-\infty,\infty)$ and is not periodic, so the spacetime is actually better thought of as AdS$_3$ presented in Rindler-AdS$_3$ coordinates, with compactified Rindler time $\phi$. Then there are CTCs in the asymptotic region; see Fig.~\ref{fig:penrose}. 

At the other side of the Rindler horizon at $\tilde{r}=r_0$, the coordinate $\phi$ becomes timelike, so this is a chronology horizon. This metric and the one in the lower patch \eqref{BTZ1} are seen to be the same by relating $\tilde{r}^2=r_0^2-r^2$. Travelling between the two asymptotic regions involves crossing two different horizons. Starting from \eqref{BTZ1}, we first cross the black hole horizon of the vector $\partial_t$ at $r=r_0$ into the BTZ interior, where $\phi$ and $t$ are spacelike and $r$ is timelike. From here, we cross the horizon of $\partial_\phi$, which is a chronology horizon---at $r=0$ in \eqref{BTZ1} or equivalently at $\tilde{r}=r_0$ in \eqref{Rads}---and emerge in the exterior of Rindler-AdS$_3$, with periodic timelike $\phi$ and spacelike $t$ and $\tilde{r}$.

Therefore, the geometries for the lower and upper patches are simply different regions of the extended BTZ manifold, which lie on different sides of the bulk chronology horizons.

This connection between the two patches is somewhat obscured when working with the coordinate $z$. In this Fefferman-Graham gauge, there appear `holes' of the kind we encountered in \eqref{tohole}. In the present case,  when $\alpha<0$ ($r_0^2>1$) the lower-patch BTZ has a gauge hole (a circle) inside the BTZ horizon, at $r^2=r_0^2-1$; in the upper patch, when $0<\alpha<1/2$  ($r_0^2<1$) the hole is outside the Rindler horizon, at $\tilde{r}=1$. There is only one instance of $\alpha<1/2$ solutions where the Fefferman-Graham gauge covers all the geometry, namely, the state with $\alpha=0$, $r_0=1$. 

In these geometries the parameters $\alpha$ and $\Delta$ need not bear any relation to each other. While the conventional boundary geometry for BTZ has no parameter independently of the black hole mass, the choice of a Misner-AdS$_2$ boundary introduces a new parameter. However, while this is true of the Lorentzian sections of the geometry, it need not hold in their Euclidean continuations. If we continue the time coordinate $\phi$ of the upper patch to imaginary values, it is straightforward to verify that Euclidean regularity demands that $r_0=2\pi/\Delta$, that is,
\beq\label{alphaDelta2}
\alpha=\frac12-\frac{2\pi^2}{\Delta^2}\,.
\eeq
If the quantum state is to be regular throughout the complex time ($\phi$) plane (to the extent that it can be well defined in a background with CTCs), then this condition should be satisfied. 

\paragraph{Traversable bulk wormhole.}

It is clear from our analysis above that all the solutions with $\alpha<1/2$ share the same qualitative physical features---they are all described by the extended static BTZ manifold. The solution with $\alpha=0$ is the only one completely covered in the Fefferman-Graham gauge, so we shall examine it more closely in order to illustrate the nature of states with $\alpha<1/2$.
 
Let us write the $\alpha=0$ bulk geometry \eqref{AdStimemach} using the proper distance (Gaussian) coordinate $\sigma=\ln(2/z)$. We also revert to the Misner $\psi$ coordinate using \eqref{AdStimemach}, since this allows to maximally cover Misner-AdS$_2$. Then \eqref{AdStimemach} becomes
\begin{align}\label{emptybulk}
ds^2
= d\sigma^2 + \cosh^2\sigma\lp -(d\tau+\tau d\psi)^2+d\psi^2\rp
\end{align}
with $\psi\sim\psi+\Delta$.

The stress tensor of this state only contains the trace anomaly, since as we said, the thermal and Casimir components exactly cancel each other. This cancellation is a property peculiar to two-dimensional CFTs, which makes it possible that the CFT is unexcited even if the spacetime evolves dynamically.

If $\psi$ ranged from $-\infty$ to $\infty$, the metric \eqref{emptybulk} would just describe global AdS$_3$ foliated by AdS$_2$ slices. But since $\psi$ is periodic, and is a boost coordinate, we obtain a BTZ black hole, extended across its future and past chronological `singularities'. This was shown in our previous analysis, but we can directly verify that the change
\begin{equation}
    r=\sqrt{|1-\tau^2|}\cosh\sigma\,,\qquad t=\arctanh\lp\frac{\tau}{\tanh\sigma}\rp
\end{equation}
brings \eqref{emptybulk} into the BTZ metric \eqref{BTZ1} for $\tau^2<1$, and into Rindler-AdS$_3$ \eqref{Rads} for $\tau^2>1$, in both cases with  $r_0=1$. Then, even though the form of \eqref{emptybulk} does not make it apparent, it has a black hole horizon at
\begin{equation}
    \tau^2=\tanh^2\sigma\,.
\end{equation}
The exterior of the black hole is the region $\tau^2<\tanh^2\sigma$, which lies entirely within the non-CTC part $\tau^2<1$, while for times $\tanh^2\sigma<\tau^2<1$, we are in the black hole interior. Thus, CFT excitations of the boundary will, in the dual bulk description, enter a black hole horizon at a time depending on their energy scale.

Using these coordinates, with $-\infty<\tau<\infty$ and $0\leq \psi\leq \Delta$, the metric \eqref{emptybulk} covers the conventional BTZ black hole geometry plus two Rindler-AdS$_3$ regions, one to the future and another to the past of the BTZ chronology horizons at $\tau=\pm 1$. 
This is not the maximally extended manifold, but still it covers a larger portion than the conventional BTZ geometry. The boundary of BTZ is conformal to only the region $\tau^2<1$ of Misner-AdS$_2$.

What is, then, the interpretation of this  state? The bulk geometry \eqref{emptybulk} describes a traversable wormhole connecting two separate asymptotic regions at $\sigma\to +\infty$ and $\sigma\to -\infty$, connected through the bulk across Misner chronology horizons at $\tau^2=1$.\footnote{This bulk wormhole is completely unrelated to any wormhole that the Misner-AdS$_2$ spacetime may have originated from (see Sec.~\ref{sec:wormhole}).} Strictly speaking, it is a wormhole only during the time that CTCs are not present, when the spatial topology of the bulk is that of a cylindrical wormhole, $\R_\sigma\times S^1_\psi$. Afterwards, the spatial topology changes to $\R_\sigma\times \R_t$. This provides a simple example of the theorem that relates spatial topology change to the presence of CTCs \cite{Geroch:1967fs}. 

The fact that we have managed to construct a traversable 3D wormhole without introducing any negative energies in the bulk should give us pause. This bulk is locally AdS$_3$ without any matter in it, but the results that forbid traversable wormholes are evaded by having CTCs. It is impossible to cross the wormhole while remaining within a spacetime region that is free of CTCs. Since one then goes through a change in the spatial topology, one might also say that the wormhole becomes traversable by unwrapping its compact circle. The two boundaries then become spatially infinite, and are connected through the endpoints at $t=\pm \infty$.

The same qualitative conclusions can easily be seen to apply to all the bulk geometries with $\alpha<1/2$. Next we turn to their dual CFT interpretation.

\paragraph{Entangled time machines.} Since the bulk geometry has two different boundaries connected by a wormhole, we have two copies of the CFT in separate but identical Misner-AdS$_2$ spacetimes. Then, a state of the CFT with $\alpha<1/2$ is entangled with another copy of the same CFT. The entanglement is responsible for what we have referred to as the thermal component of the stress-energy tensor. The state might then seem similar to a thermofield double state. However, the static thermofield double is dual to a non-traversable wormhole, where the entanglement cannot be used to send information between the two CFTs. In our entangled time machines, it would seem like the time dependence of the Misner-AdS$_2$ geometries, where the CFTs live, makes the initial $(\tau=0)$ thermofield double state evolve to enable teleportation between the two systems. 

\paragraph{Does entanglement enable chronology violation?}

This suggests that by entangling two time machines we can evade the split-bulk chronology protection. The entanglement generates a bulk wormhole through which CFT excitations can cross the bulk chronology horizon.\footnote{What we saw for Misner-AdS$_2$ in Sec.~\ref{sec:motion} also applies here: although there exist null geodesics that end at the chronology horizon, there are others which cross it smoothly. Generic timelike geodesics also cross it in finite proper time.} Note the subtlety of the mechanism: from the boundary viewpoint, the field excitation is not crossing to the other side of the chronology horizon where the CFT diverges. Instead, by diving inside the bulk, the field excitation reappears in the CTC region \emph{of a different spacetime}. Although not a conventional violation of chronology, the possibility to enter a region with CTCs looks worrisome.

However, we think this is not a viable physical mechanism, since 
the traversability of the bulk wormhole is likely an artifact of the large $N$ (large $c$) limit of the holographic CFT. In \cite{Emparan:2021yon} we will present arguments, similar to those in \cite{Emparan:2020rnp}, that corrections in $1/c$, corresponding to quantum effects in the 3D bulk, turn the chronology horizon into a strong, spatial curvature singularity. This should close up the passage into the CTC region, and break the dual entanglement between the CFTs. Since the argument does not rely on the divergence of the holographic stress tensor, it also applies to the zero-stress $\alpha=0$ state.

Thus, chronology protection is again upheld without appealing to gravitational backreaction on the Misner-AdS$_2$ geometry. Self-interaction of the quantum fields, now beyond the leading planar approximation, is enough to banish any communication between different sides of the chronology horizons.\footnote{Observe that a signal can be sent through the bulk wormhole from the non-CTC region of one CFT to the CTC region of the other CFT without violating the no-transmission principle of \cite{Engelhardt:2015gla}, since the CFTs in these regions are entangled with each other. However, this principle would be violated in the maximally extended BTZ manifold; it is then fitting that $1/c$ corrections will restore the system back to sanity much like in \cite{Emparan:2020rnp}.}

\subsection{$\alpha=1/2$: massless BTZ}

We will briefly mention now the borderline case $\alpha=1/2$. It is easily understood as the limit $r_0=0$, $\alpha\to 1/2^-$, of the wormhole geometries. The bulk is an extension of the massless BTZ solution. It is then possible to cross to the CTC side, where the geometry appears as Poincar\'e-AdS$_3$ with periodically identified time.

\section{Bulk duals for flat Misner spacetime}
\label{bulk of flat misner}

We have seen in Sec.~\ref{subsec:toflat} that we can recover the flat Misner spacetime as the near-chronology-horizon limit of Misner-AdS$_2$. We can apply the same limit to the CFT stress tensor. The anomalous trace disappears in flat space, but the traceless component of the stress tensor retains the form \eqref{stressalpha} in null coordinates. Free conformal scalars in Misner spacetime were first studied in \cite{Hiscock:1982vq}.

We could now apply the bulk reconstruction procedure to this CFT state, but it is also straightforward to carry over all our results for Misner-AdS, taking the limit of the boundary coordinates and simultaneously zooming in on small values of $z$ doing
\beq\label{zflat}
z\to \sqrt{\epsilon}\, z\,.
\eeq
Either way, we find the bulk metric 
\begin{align}\label{flattimemachznull}
ds^2
&=\frac{1}{z^2}\Biggl[dz^2 - \lp 1 +\frac{\alpha^2}{4(X^+ X^-)^2} z^4\rp dX^+ dX^- -\frac{z^2}{2}\alpha\lp \lp \frac{dX^+}{X^+}\rp^2+ \lp  \frac{dX^-}{X^-}\rp^2\rp\Biggr]\\
&=\frac{1}{z^2}\Biggl[dz^2 + \lp 1 -\frac{\alpha z^2}{4T}\rp^2 \frac{dT^2}{2T} -\lp 1 +\frac{\alpha z^2}{4T}\rp^2 2T d\phi^2\Biggr]\,.\label{flatztauphi}
\end{align}
The identifications are made along the orbits of the bulk Killing vector
\beq
k=\frac{\partial}{\partial\phi}=X^+\frac{\partial}{\partial X^+} - X^-\frac{\partial}{\partial X^-} \,,
\eeq
with periodicity $\Delta$. The action of the vector has fixed points at
\beq
z=\sqrt{\frac{2}{\alpha} X^+ X^-}=2\sqrt{\frac{-T}{\alpha}}
\eeq
and the absence of conical singularities at these points requires that $\alpha$ is determined in terms of $\Delta$ by the same value \eqref{alphaDelta} as before. These fixed points appear only in the region where $X^+ X^->0$, $T<0$, where there are no CTCs.

As before, these metrics are locally equivalent to AdS$_3$, and it is straightforward to find the coordinate changes from them to more conventional coordinates $(t,r,\phi)$, namely\footnote{E.g., take the flat limit of \eqref{rtauz}, where the coordinate $r$ must not be scaled by $\epsilon$. It is also easy to directly find maps to Poincar\'e AdS$_3$, see App.~\ref{app:Pmap}.}
\begin{align}
r^2&= \frac{2|T|}{z^2}\lp 1+ \frac{\alpha}{4}\frac{z^2}{T}\rp^2
\,,\\
t&=\frac1{2}\ln \left|\frac{2}{T} 
\lp \frac{8T+z^2\lp \sqrt{1-2\alpha}-1\rp^2}{8T+z^2\lp \sqrt{1-2\alpha}+1\rp^2}\rp^{\frac1{\sqrt{1-2\alpha}}}
\right|\,.
\end{align}
These then reproduce the same bulk duals \eqref{gads3} as for Misner-AdS$_2$: global AdS$_3$ for $\alpha>1/2$---complete in the lower patch where $T<0$, conically singular in the upper one where $T>0$---while for $\alpha<1/2$ we get again the BTZ solution in the lower patch, extended to periodic-time Rindler-AdS$_3$ in the upper patch.

The zero-stress state $\alpha=0$, obtained either from \eqref{flatztauphi} or as the flat-Misner limit of \eqref{emptybulk}, can be written very simply as 
\beq\label{flatmisnerbulk}
ds^2=\frac1{z^2}\lp dz^2 -2 d\psi\,( dT+Td\psi)\rp\,,
\eeq
with periodic $\psi$. Despite appearances, this is not the massless BTZ black hole, since the global identifications are those of a positive mass black hole. Thus, rather than a zero-temperature configuration, this state should be viewed as having a thermal component of temperature $2\pi/\Delta$ that cancels the negative Casimir energy.

\section{Zero-stress states in Misner spacetimes?}\label{sec:zerostress}

From the bulk viewpoint, the states with zero stress tensor do not seem to differ in any special way from other states with $\alpha<1/2$. Still, it is peculiar that the CFT does not show any divergence on the chronology horizon. 

Similar zero-stress states have been considered for free conformal fields in Misner spacetimes---flat and AdS Misner spacetimes \cite{Krasnikov:1995jn,Sushkov:1995hg,Li_1998,Li_1999}. In refs.~\cite{Li_1998,Li_1999} they were referred to as `self-consistent', since, having zero stress tensor, they (trivially) solve the backreaction equations. However, it has been disputed whether they are physical states. It has been argued that, despite the vanishing stress tensor, the state of the field is singular on the chronology horizon \cite{Cramer:1996vz,Cramer_1998} and may be unstable to the inclusion of interactions \cite{Hiscock:2000jq}. And moreover, the cancellation of the stress tensor requires to fine-tune the periodicity to $\Delta=2\pi$.

Are our holographic zero-stress states related to the ones found for free theories? The geometries of our holographic bulks do not appear to exhibit any qualitative differences around the value $\alpha=0$, and furthermore, the states that we have found exist, as Lorentzian solutions, for all values of $\Delta$. However, Euclidean regularity demands that the state for $\alpha=0$ must have $\Delta=2\pi$, see \eqref{alphaDelta2}. So this may connect them to the states in \cite{Li_1998,Li_1999}. The finding in \cite{Hiscock:2000jq} that perturbative interactions render the state singular would seemingly disagree with our strong-coupling results. However, as we have argued, the smoothness of the state is not expected to survive $1/c$ corrections of the CFT, regardless of the vanishing of the leading-order holographic stress tensor. So our zero-stress states may indeed be strong coupling counterparts of those in \cite{Li_1998,Li_1999}.



\section{Time machines from wormholes}
\label{sec:wormhole}

As we mentioned in the introduction, time machines seem to be a generic feature of traversable wormholes. We will show in \cite{future} that this conclusion is premature, but the connection gives a physical motivation for studying their properties, so we will explain it here. We follow previous literature \cite{PhysRevD.43.3878}, but highlight the connection to the Misner-AdS spacetime which, to our knowledge, has not been made before.

The geometry of a spherically symmetric wormhole can be written in the form
\beq\label{sphwh}
ds^2=-e^{-2\Phi(\lambda)}dt^2+d\lambda^2+R^2(\lambda)d\Omega_2\,,
\eeq
where we take $R(\lambda)$ to be a U-shaped function of $\lambda\in(-\infty,+\infty)$, with $R^2\to \lambda^2$ as $|\lambda|\to\infty$. The mouths are where the behavior $R\simeq |\lambda|$ sets in. In \eqref{sphwh} the two mouths live in different asymptotic regions, but we will want them connected within the same space, so that it is possible travel between them through the `exterior' universe. In an asymptotically flat universe, this necessarily breaks the spherical symmetry of \eqref{sphwh}, but we will still consider it to be a good approximation near each of the mouths. We do not assume that the tube is short, neither compared to its characteristic radius nor to the exterior distance between mouths.

The gravitational potential $\Phi(\lambda)$ is finite everywhere, so there are no horizons and the wormhole is traversable. We assume that $\Phi(\lambda)\to\Phi_\infty^\pm$ as $\lambda\to \pm\infty$. Typically, these asymptotic values are reached not far from the mouths. 

For our present discussion, we will not specify the gravitational theory and the matter that support this geometry. But we must bear in mind that having a self-consistent model of a wormhole is crucial for deciding what constraints exist on the features, such as the length, of a physically allowed wormhole. It is also necessary for examining its stability and its reaction to attempts to turn it into a time machine, as we will do in \cite{future}.

In order to create a time machine, we set the two mouths at different gravitational potentials, $\Phi_\infty^+\neq \Phi_\infty^-$, e.g., by placing a heavy object near one of the mouths. Then, a time-shift will grow between the two sides. A particle traveling in a circular trajectory along the wormhole will experience a net gravitational force and will get accelerated, gaining or losing energy along the way, so this is a non-potential gravitational field. 

Say that, before introducing any redshift between the two mouths, $L_\textrm{wh}$ is the asymptotic exterior time that a light ray takes to cross the wormhole tube joining two chosen points near each of the mouths, and $d_\textrm{out}$ is the time in which a light ray travels between the same two points across the outside universe. Then, a chronology horizon appears when the redshifting effect has been operating for an asymptotic time\footnote{For simplicity we are assuming a small redshifting effect, so that $e^{\Phi_\infty^+ - \Phi_\infty^-}\simeq 1 + \Phi_\infty^+ - \Phi_\infty^-$.}
\beq\label{TCH}
T_\textrm{CH}=\frac{L_\textrm{wh}+d_\textrm{out}}{|\Phi_\infty^+ - \Phi_\infty^-|}\,.
\eeq
 At any later time, a time machine is present. Notice that this does not require that the wormhole we started with is short, in the sense that $L_{wh}<d_{ext}$. If we begin with a very long wormhole, we simply have to wait a correspondingly long time to form the time machine. Of course, as the redshift operates, the wormhole becomes shorter in the sense that the time to cross the wormhole, measured by an exterior observer, gets smaller (in one direction -- in the other it gets bigger). But, to the extent that one can make a distinction between exterior and interior,\footnote{In general, such a distinction is not precisely defined.} the proper length of the wormhole tube can still remain much longer than the outside separation between the mouths. 

An important assumption in this argument is that, as we mentioned before, the wormhole geometry remains fixed throughout, and in particular is not affected by the quantum energy fluxes induced by the time-dependent geometry (which we have computed), even when these become large. In this article we will not dwell further on this matter.

\subsection*{Wormhole time machine as Misner-AdS}

We now make a simplification and study the $1+1$ geometry seen by a particle, or a field excitation, traveling in a circle along the wormhole. Such a restriction is well justified when the field excitation is constrained to move along such lines, as is the case in the magnetic-line model of \cite{malda} or in the cosmic string model of \cite{Fu:2019vco}. In other cases, the model may miss relevant features due to propagation in directions transverse to these worldlines \cite{Visser:1995cc}, but, at any rate, the truncation allows to develop a useful `standard wormhole time machine model'. Fixing the angle on the sphere in \eqref{sphwh} gives
\beq\label{2dtimemac}
ds^2=-e^{-2\Phi(\lambda)}dt^2+d\lambda^2\,,
\eeq
where points along $\lambda$ are periodically identified, $\lambda\sim \lambda +L_\lambda$,  with $L_\lambda$ the total length of the wormhole circle. For a typical wormhole, the redshift $\phi$ grows in the interior of the throat, reaching a maximum at its middle. 

To make a time machine, we assume that $\Phi(0)\neq \Phi(L_\lambda)$. The time coordinates $t$ at $\lambda=0$ and $t'$ at $\lambda=L_\lambda$ must then be related by
\beq\label{timeident}
e^{-\Phi(L_\lambda)}t'=e^{-\Phi(0)}t\,,
\eeq
or equivalently, we identify the spacetime points
\beq
(t,0)\sim (e^{-\Phi(0)+\Phi(L_\lambda)}t,L_\lambda)\,.
\eeq
%
The spacetime \eqref{2dtimemac} can always be Weyl-transformed to the form
\beq\label{premisner}
ds^2=-e^{-2a\psi}dt^2+d\psi^2\,,
\eeq
with constant $a$, by a conformal change of coordinates $\lambda\to \psi$ that, importantly, does not involve $t$. This transformation uniformizes the path, making the redshift between the mouths grow linearly in time. The new geometry, which \cite{PhysRevD.43.3878} referred to as the `standard model', is locally the same as the Poincar\'e metric in AdS$_2$ with curvature radius $1/a$, but now with points identified in such a way that
\beq
(t,\psi)\sim (e^{a L} t,\psi + L)\,,
\eeq
where $L$ is the new periodicity of $\psi$. The coordinate $t$ here is actually the same as in \eqref{pads2}. Then, if we change
\beq
t = \tau e^{a\psi}
\eeq
the geometry \eqref{premisner} becomes the same, locally and globally, as the Misner-AdS$_2$ spacetime of \eqref{misnerads2}. The appearance of CTCs in this model is in direct correspondence with that in the original wormhole, \eqref{TCH}.

\section{Discussion}
\label{sec:final}

We have solved a CFT on a time machine background by constructing its bulk dual. 
The analysis of holographic quantum fields in this spacetime not only leads to a divergent stress tensor at the chronology horizon, like that found for free scalars in \cite{PhysRevD.43.3878}; it also reveals that the holographic bulk dual geometry implements in a simple, appealing manner the impossibility that quantum excitations cross to the side with CTCs. The bulk is separated into two patches, with and without CTCs, but without any curvature singularity. The lower (no-CTC) spacetime is a geodesically complete, global AdS$_3$ geometry totally disconnected from the upper (CTC) region; they are only connected through the boundary where the chronology horizon lies. Such a separation clearly indicates that no field excitation can ever cross to the region with CTCs. From the CFT viewpoint, the effect is entirely due to the self-interaction of the strongly coupled fields. 


We have also identified a class of bulk geometries that describe entangled states of CFTs in two separate time machine spacetimes. These bulk wormholes connect the two time machines in a way that suggests that entanglement may allow to alter the normal chronological order. However, we expect (and in \cite{Emparan:2021yon} will further argue) that, again, the self-interaction of the quantum fields protects the chronology: Traversability of the bulk wormhole is a leading large-$N$ effect that will not survive higher order corrections.

One might wonder how we can say anything at all about the upper patch beyond the chronology horizon, given that a divergent stress tensor develops beforehand. Indeed, it was emphasized in \cite{Dias:2019ery} that one cannot make meaningful predictions about the state of the quantum fields beyond a Cauchy horizon (nor about the spacetime itself)\footnote{Note that we argued in \cite{Emparan:2020rnp} that the BTZ black hole does not violate the strong cosmic censorship once we include sufficiently high loop corrections to the spacetime.}. Simply put, we cannot pose an initial value problem on a singular slice, although we can put initial values on a slice just beyond it. However, these initial values can be chosen completely independently of the state before the Cauchy horizon.  In the bulk construction, this is manifest since the lower-patch bulk is geodesically complete and therefore cannot affect anything ``outside'' its domain of dependence. 
Nevertheless, we can certainly discuss the possible states of the CFT in the upper patch and their bulk duals; these states might not make sense beyond a singular Cauchy horizon, but they can be analyzed as independently prepared states with no ties to the lower patch, as we have done in Sec.~\ref{sec:bulk}. 


Our result that self-interaction makes it impossible for CFT excitations to cross the chronology horizon can be expected to apply also to any other non-gravitational system that couples to the conformal fields without greatly changing the state of the field. Then, that system can be holographically mapped as propagating in a complete bulk, and never crossing the chronology horizon. The interaction with the CFT makes its energy grow unbounded, but in the dual bulk its proper motion never reaches any singularity. 

However, the most universal way to transfer the chronology protection to any other system is by coupling the CFT to gravity. The idea that a strong gravitational backreaction should render the chronology horizon an impassable curvature singularity was proposed in \cite{PhysRevD.46.603, Morris:1988tu, Visser:1995cc, PhysRevD.43.3878}, but it remained inconclusive since the backreaction could not be consistently computed. The methods we have developed will be extended in \cite{Emparan:2021yon} to investigate this problem in the same class of models as in this article, confirming that complete chronology protection is achieved.


Still, we feel that a catastrophically strong backreaction should be the last resort for upholding the normal chronological order, specially in more realistic setups such as the asymptotically flat wormholes of Sec.~\ref{sec:wormhole}. We expect that low-energy gravitational backreaction can do the job in a more controllable manner, but, possibly, still equally general. As we will explain in \cite{future}, in a wormhole time machine the achronal averaged null energy condition (AANEC), which is supposed to be satisfied by physical quantum matter (but which plays no role in the two-dimensional models in this article), implies that significant, but controllable, changes in the spacetime must occur well before the chronological horizon is reached, which will prevent the appearance of CTCs.

This goes beyond the statement (following the theorem in \cite{Graham_2007}) that ``the AANEC directly forbids time machines''. Declaring this leaves us still in the dark about how we will fail if we try to turn a wormhole into a time machine as in Sec.~\ref{sec:wormhole}.\footnote{It is, in this regard, similar to the difference between \emph{stating} that reaching superluminal speeds is forbidden by the postulates of special relativity, and \emph{explaining} what happens when we try to accelerate a body beyond light speed.} In particular, it does not tell us what kind of physics is responsible for the failure: is it eventual Planck-scale physics, as proposed in \cite{PhysRevD.46.603, PhysRevD.43.3878, Morris:1988tu, Visser:1995cc, Kay_1997, Cramer_1998}, or is it instead other, more controllable, low energy physics? If, as we shall argue in  \cite{future}, it is the latter, then chronology may be protected without dramas in the geometry.


\section*{Acknowledgments}
We thank Tom\'as Andrade, Evan Coleman, and Mikel S\'anchez-Garitaonandia for useful discussions. Work supported by ERC Advanced Grant GravBHs-692951, MICINN grants FPA2016-76005-C2-2-P and PID2019-105614GB-C22, and AGAUR grant SGR-2017-754. We also acknowledge financial
support from the State Agency for Research of the Spanish Ministry of Science and Innovation
through the “Unit of Excellence Mar\'{i}a de Maeztu 2020-2023” award to the Institute of Cosmos Sciences (CEX2019-000918-M).
\appendix

\section{Hyperboloidal embeddings}
\label{app:B}

In this section, we rewrite our main AdS$_2$ metrics in terms of the embedding in a three-dimensional space; such transformations are useful when discussing different patches of the manifold.

The Misner-AdS$_2$ metric \eqref{misnerads2} (with $a=1$) can be obtained from the hyperboloid
\beq
-V^2-W^2+X^2=-1\,\quad \text{in}\quad ds^2=-dV^2-dW^2+dX^2 \,,
\eeq
by making
\beqa
V+X&=&e^{-\psi}\,,\nn\\
V-X&=&e^{\psi}(1-\tau^2)\,,\nn\\
W&=&\tau\,.
\eeqa
These coordinates only cover $V+X>0$ and therefore there exist incomplete geodesics that cannot be extended into $V+X<0$. In Sec.~\ref{sec:motion}  this is described in $(\tau,\psi)$ coordinates.

The identification $\psi\sim\psi+\Delta$ that makes Misner-AdS$_2$ into a time machine is 
\beq
V\pm X\sim e^{\mp \Delta}(V\pm X)\,,
\eeq
which is an AdS$_2$ boost. This spacetime can then also be obtained by going to Rindler coordinates and identifying along the boost. Maintaining the $\tau$ coordinate and changing $\psi$ to another periodic coordinate $\phi$, such that
\beq
V\pm X=\sqrt{1-\tau^2}e^{\mp \phi}\,,\qquad W=\tau\,,
\eeq
we recover the metric \eqref{tauphimetric} for $\tau^2<1$. For $\tau^2>1$, we use, instead,
\beq
V\pm X=\pm\sqrt{\tau^2-1}e^{\mp \phi}\,,\qquad W=\tau\,.
\eeq

\section{Bulk reconstruction in AdS$_3$}
\label{app:C}

We will follow \cite{de_Haro_2001}, where it is shown that in an AdS$_3$ bulk the Fefferman-Graham expansion truncates exactly to the form
\beq\label{ads3rec}
ds^2=\frac{\ell^2}{z^2}\lp dz^2 + \lp g_{ij} + z^2 g_{ij}^{(2)}+\frac{z^4}4g^{(2)}{}_i{}^k g_{kj}^{(2)}\rp dx^i dx^j\rp\,.
\eeq
Indices are raised and lowered with the boundary metric $g_{ij}$ (we omit the index $(0)$). The term $g_{ij}^{(2)}$ determines the stress tensor as
\beq
g_{ij}^{(2)}=\frac12 \lp t_{ij}-R g_{ij}\rp\,,
\eeq
where $R$ is the scalar curvature of $g_{ij}$ and $t_{ij}$ is the `geometric stress tensor', which satisfies
\beq
\nabla_i t^{ij}=0\,,\qquad t^i{}_i =R\,,
\eeq
and yields the physical holographic stress tensor as
\beq\label{Tt}
T_{ij}=\frac{\ell}{16\pi G}t_{ij}=\frac{c}{24\pi}t_{ij}\,.
\eeq
It is often convenient to separate its traceless part,
\beq
\hat{t}_{ij}=t_{ij}-\frac12 R g_{ij}\,,
\eeq
in terms of which
\beq
g_{ij}^{(2)}=\frac12 \lp \hat{t}_{ij}-\frac12 R g_{ij}\rp\,.
\eeq

Now we insert this into \eqref{ads3rec} to obtain
\beq
ds^2=\frac{\ell^2}{z^2}\left[ dz^2 + \left[\lp \lp 1-\frac{R}{8}z^2\rp^2+\frac{\hat{t}^2}{32}z^4\rp g_{ij}
+\frac{z^2}{2}\lp 1-\frac{R}{8}z^2\rp\hat{t}_{ij}\right] dx^i dx^j\right]\,,
\eeq
where we denote
\beq
\hat{t}^2\equiv \hat{t}^{ij}\hat{t}_{ij}\,,
\eeq
and we have used that
\beq
\hat{t}_i{}^k\hat{t}_{kj}=\frac{\hat{t}^2}2 g_{ij}\,.
\eeq

Now consider the case where the boundary metric is written in null coordinates $x^\pm =t\pm x$ as
\beq
g_{ij}dx^i dx^j = -g(x^+,x^-)dx^+ dx^-
\eeq
so that $g_{+-}=-g/2$, $g^{+-}=-2/g$, and
\beq
\hat{t}_{ij}dx^i dx^j =\hat{t}_+ \lp dx^+\rp^2+\hat{t}_- \lp dx^-\rp^2\,,
\eeq
\beq
\hat{t}^2=\frac{8}{g^2}\hat{t}_+\hat{t}_-\,.
\eeq
Then
\beq
\begin{split}
ds^2 =  \frac{\ell^2}{z^2}\Biggl[& dz^2 - \lp \lp 1-\frac{R}{8}z^2\rp^2g+\frac{\hat{t}_+\hat{t}_-}{ 4g}z^4\rp dx^+ dx^-  \\
& +\frac{z^2}{2}\lp 1-\frac{R}{8}z^2\rp\lp\hat{t}_+ \lp dx^+\rp^2+\hat{t}_- \lp dx^-\rp^2\rp\Biggr]\,.
\end{split}\label{recons}
\eeq

The BTZ black hole is recovered for $g_{ij}=\eta_{ij}$ (so $R=0$) and constant $\hat{t}_{tt}=\hat{t}_{xx}= \mu$; in null coordinates, this is $g=1$, and $\hat{t}_+=\hat{t}_-=\mu/2$. The dimensionless constant is $\mu=8GM$. For $\mu=-1$ we recover global AdS$_3$. 

Apply this to the Misner-AdS$_2$ spacetime, which as we have seen in Sec.~\ref{subsec:misnerads} is a quotient of the Poincar\'e AdS$_2$ geometry, with 
\beq
g=\frac{4}{(x^+-x^-)^2}
\eeq
(so that $R=-2$), and with the traceless stress tensor \eqref{stressalpha},
\beq\label{stressalphat}
\hat{t}_\pm =-\frac{\alpha}{(x^\pm)^2}
\eeq
which is conserved when $\alpha$ is constant. Substituting in \eqref{recons} we obtain the bulk metric \eqref{AdStimemach}.
If, instead, we take $g=1$ with the same stress tensor, we find  \eqref{flattimemachznull}.

\section{Mapping the flat-Misner bulk to Poincar\'e AdS$_3$}
\label{app:Pmap}

Let us see how the spacetime \eqref{flattimemachznull} can be obtained by a coordinate transformation from Poincar\'e AdS$_3$,
\beq\label{PAdS3}
ds^2=\frac{\ell^2}{\tilde{z}^2}\lp d\tilde{z}^2-dy^+ dy^-\rp\,.
\eeq

Observe first that, in the boundary theory, the change
\beq\label{yxgamma}
y^\pm=(X^\pm)^\gamma
\eeq
generates, through the Schwarzian, 
\begin{equation}
\hat{t}_\pm=-\left\{y^\pm,X^\pm\right\}\,,
\end{equation}
the stress tensor \eqref{stressalphat} with 
\beq\label{alphagamma}
\alpha=\frac{1-\gamma^2}{2}\,.
\eeq
Comparing to \eqref{alphaDelta2}, we expect that
\begin{equation}
    \gamma=\frac{2\pi}{\Delta}\,.
\end{equation}
The bulk change of coordinates that reproduces this transformation can be obtained from, e.g., \cite{Roberts:2012aq}, to find
\begin{align}
y^\pm=&(X^\pm)^\gamma\,\frac{X^+X^- +(\gamma^2-1)z^2/4}{X^+ X^--(\gamma-1)^2 z^2/4}\,,\nn\\
\tilde{z}=&z\, \frac{\gamma(X^+ X^-)^{\frac{\gamma+1}2}}{X^+ X^--(\gamma-1)^2 z^2/4}\,.
\end{align}

When $\gamma=0$ we get an invalid coordinate change, but this can be corrected by taking an appropriate limit. Then, the change of coordinates,
\begin{align}\label{coordchange}
y^\pm=&\ln x^\pm- \frac{z^2/2}{X^+ X^-- z^2/4}\,,\nn\\
\tilde{z}=&z\, \frac{\sqrt{X^+ X^-}}{X^+ X^-- z^2/4}
\end{align}
gives \eqref{flattimemachznull} with $\alpha=1/2$, which is indeed the value that obtains from the Schwarzian stress tensor in the boundary when changing $y^\pm=\ln X^\pm$. 

This change of coordinates can be obtained as a limit of the ones we introduced for Misner-AdS$_2$.

\newpage

\bibliography{bibi}
\bibliographystyle{utcaps}

\end{document}